\documentclass[prl,twocolumn,showpacs]{revtex4}

\include{epsf}

\begin{document}

\title {Macrodimers: ultralong range Rydberg molecules}

\author{Christophe Boisseau}
\author{Ionel Simbotin}
\author{Robin C\^{o}t\'{e}}\email{rcote@phys.uconn.edu}

\affiliation{Physics Department, University of Connecticut, 
2152 Hillside Rd., Storrs, Connecticut 06269-3046}

\date{\today}

\begin{abstract}

We study long range interactions between two Rydberg atoms
and predict the existence of ultralong range Rydberg
dimers with equilibrium distances of many thousand Bohr radii. 
We calculate the dispersion coefficients $C_{5}$, $C_{6}$ and 
$C_{8}$ for two rubidium atoms in the same excited level $np$,
and find that they scale like $n^{8}$, $n^{11}$ and 
$n^{15}$, respectively. We show that for certain molecular symmetries,
these coefficients lead to long range potential wells that can 
support molecular bound levels. Such macrodimers would be very sensitive
to their environment, and could probe weak interactions.
We suggest experiments to detect these macrodimers.

\end{abstract}


\maketitle

New techniques used for cooling and trapping of atoms \cite{atoms} 
and molecules \cite{molecules}, and which led to the realization 
of atomic Bose--Einstein condensation \cite{bec}, have also been 
applied to experiments with ultracold plasmas \cite{cold-plasma},
and ultracold Rydberg atoms \cite{cold-rydberg}. 
The exaggerated properties of Rydberg atoms provide a fertile
ground for new physics. For example,
transport properties of ultracold gases doped with ions
were recently explored and extended to cold Rydberg samples \cite{cold-ion},
while entangled states relevant for quantum computing can also
be produced with ultracold Rydberg atoms 
\cite{qc1}. Finally, the creation of ``trilobite" Rydberg
molecules was proposed \cite{greene}, where one atom of the
dimer remains in its ground state while the second one is excited
to a Rydberg state.

In this paper, we explore the interactions between two Rydberg
atoms. We show that long range wells supporting several bound levels
exist for certain molecular symmetries. We explore the sensitivity of 
these wells to the particular asymptotic form of the potential
long range expansion and show that their existence is
robust. We also estimate the effect of retardation as well as
the validity of the Born--Oppenheimer approximation. 
We give numerical examples for the case of 
rubidium (Rb), and discuss experimental schemes
to detect these macrodimers. These molecules could lead to 
measurements of very weak interactions, such as vacuum fluctuations,
and provide a unique tool to study quenching in ultracold collisions.

We consider two atoms each excited by one photon from their
ground state into the same Rydberg
state $np$ \cite{ns+nd}, where $n$ is the principal
quantum number. For Rb, this corresponds to the 
$5s\rightarrow np$ transition \cite{note:photoassociation}.
At large separation $R$, 
the potential energy between two atoms can be expanded in powers of 
$1/R$ \cite{alex-review}. For two identical
atoms in the same $np$ state, it takes
the form \cite{np+np}
\begin{equation}
 V(R) = -\frac{C_{5}}{R^{5}} - \frac{C_{6}}{R^{6}} 
        - \frac{C_{8}}{R^{8}} \; ,  
\label{eq:expansion}
\end{equation}
where the dispersion coefficients $C_{5}$, $C_{6}$, and
$C_{8}$ depend on $n$. 
For the $np-np$ asymptote of the homonuclear dimers,
we have in total six pairs of degenerate molecular
states with identical coefficients for each pair: 
$^{1}\Delta_{g}$ and $^{3}\Delta_{u}$,
$^{1}\Pi_{g}$ and $^{3}\Pi_{u}$, $^{3}\Pi_{g}$ and $^{1}\Pi_{u}$,
$^{3}\Sigma^{-}_{g}$ and $^{1}\Sigma^{-}_{u}$,
and two pairs of 
$^{1}\Sigma^{+}_{g}$ and $^{3}\Sigma^{+}_{u}$ \cite{np+np}.
Note
that the degeneracy of any pair of molecular states is lifted by
the exponentially decaying exchange interactions, which can
be neglected at large enough $R$.

\begin{figure}
\centerline{\epsfxsize=3.25in\epsfbox{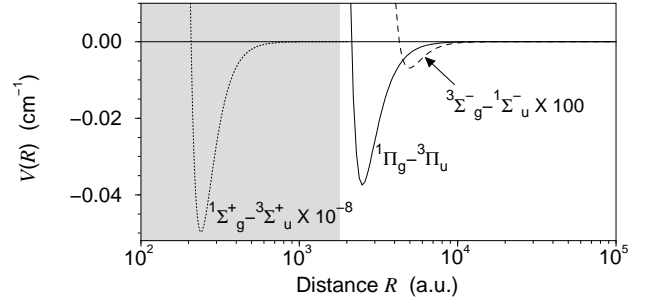}}
\caption{ Comparison of the potential curves 
	for the three pairs of states that sustain a well for
	Rb($n=20$). Inside the shaded area we have $R<R_{\rm LR}$, and
	Eq.~(\ref{eq:expansion}) is
	not adequate.}
\label{fig1}
\end{figure}

The numerical values of $C_{5}$, $C_{6}$ and $C_{8}$ \cite{note:qd}
were calculated using the expressions of Marinescu \cite{np+np}.  The
sums over the electronic states were evaluated directly, and using
the $n$-scaling of the dipole $(\propto n^{2})$, quadrupole 
$(\propto n^{4})$, and octopole $(\propto n^{6})$ matrix 
elements and the energy differences $(\propto n^{-3})$
involved, we obtained the following scaling laws: 
$C_{5}\propto n^{8}$, $C_{6}\propto n^{11}$, and 
$C_{8}\propto n^{15}$.
The magnitude and sign of the dispersion coefficients depend on the molecular
symmetry considered, and it is possible to obtain a long range
potential well with an attractive long range $R^{-5}$ contribution and
a repulsive shorter range $R^{-6}$ or $R^{-8}$ contribution (see
Fig.~\ref{fig1}).  For the $np-np$ asymptote, we found that three
pairs of degenerate molecular states give long range potential wells:
$^{1}\Pi_{g}$-$^{3}\Pi_{u}$,
$^{3}\Sigma^{-}_{g}$-$^{1}\Sigma^{-}_{u}$, and one of
$^{1}\Sigma^{+}_{g}$-$^{3}\Sigma^{+}_{u}$.

For the system to be adequately described by 
Eq.~(\ref{eq:expansion}),
the exchange energy must be negligible. To
estimate the region of validity of Eq.~(\ref{eq:expansion}), we use the
Le~Roy radius $R_{\rm LR}$ \cite{leroy} as measure of the electron
wavefunction overlap between the two atoms:  it is
given by $R_{\rm LR} = 2\left( \langle r^{2}\rangle_{A}^{1/2} + \langle
r^{2}\rangle^{1/2}_{B}\right)$, where $\langle r^{2}\rangle^{1/2}_{A,B}$ is
the rms position of the electron of atom $A$ ($B$)
\cite{note:leroy}.  If $R<R_{\rm LR}$, exchange and charge-overlap
interactions become important and Eq.~(\ref{eq:expansion}) is
not adequate to describe the system.

\begin{figure}
\centerline{\epsfxsize=3.25in\epsfbox{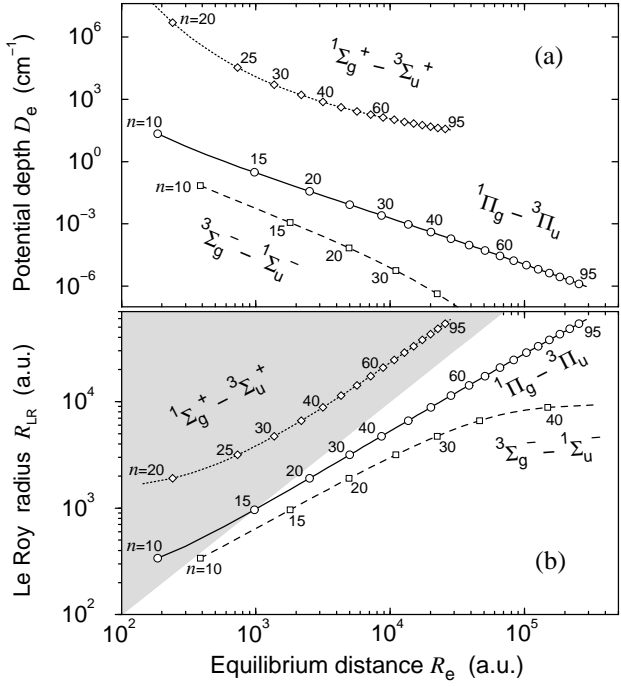}}
\caption{In (a), comparison of $D_{e}$ as a
          function of $R_{e}$ for the three
          pairs supporting long range wells, for various 
          $n$. In (b), $R_{e}$ is compared to 
          $R_{\rm LR}$ for the same states.  The shaded
          area is defined by $R_{e}<R_{\rm LR}$.}
\label{fig2}
\end{figure}

In Fig.~\ref{fig1}, we illustrate the wells for the
three pairs of degenerate states $^{1}\Pi_{g}$-$^{3}\Pi_{u}$, 
$^{3}\Sigma^{-}_{g}$-$^{1}\Sigma^{-}_{u}$,
and  $^{1}\Sigma^{+}_{g}$-$^{3}\Sigma^{+}_{u}$ of
Rb with $n=20$. The Le~Roy radius $R_{\rm LR}$ is equal to 1902 $a_{0}$,
where $a_{0}$ is a Bohr radius. The well of the pair of
states $^{1}\Sigma^{+}_{g}$-$^{3}\Sigma^{+}_{u}$ is
very deep (potential depth 
$D_{e}=4.5\times 10^{6}$ cm$^{-1}$), but is located at a
much shorter distance than the Le~Roy radius (equilibrium 
distance $R_{e}=240$ $a_{0}$): the interactions for
these states are not well described by Eq.~(\ref{eq:expansion}),
as opposed to the two other pairs. 
The well of the pair $^{1}\Pi_{g}$-$^{3}\Pi_{u}$ has a depth
$D_{e} = 3.74\times 10^{-2}$ cm$^{-1}$ and an equilibrium distance
$R_{e}= 2509$ $a_{0}$. By comparison, the well of the pair 
$^{3}\Sigma^{-}_{g}$-$^{1}\Sigma^{-}_{u}$ is much
shallower and farther away, with $D_{e}=6.88\times 10^{-5}$ 
cm$^{-1}$ and $R_{e}=4956$ $a_{0}$. In Fig.~\ref{fig2},
we compare $D_{e}$, $R_{e}$ and $R_{\rm LR}$ for the
three pairs for various values of $n$, and find the same 
general behavior. 
For the remainder of this letter, we focus
our attention on the deeper wells described by 
Eq.~(\ref{eq:expansion}), i.e., the $^{1}\Pi_{g}$-$^{3}\Pi_{u}$ pair, 
since the much shallower wells of the
$^{3}\Sigma^{-}_{g}$-$^{1}\Sigma^{-}_{u}$ pair may prove more difficult
to detect.

\begin{figure}
\centerline{\epsfxsize=3.25in\epsfbox{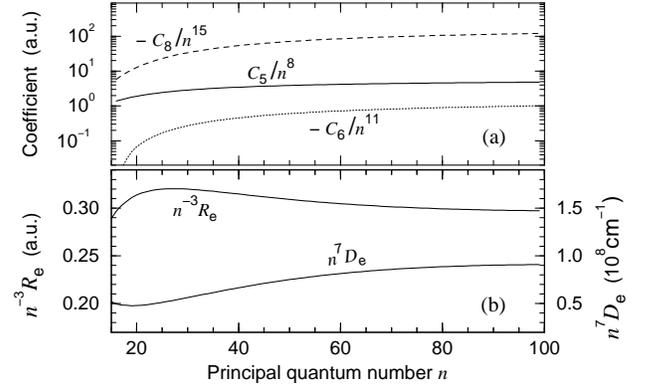}}
\caption{ Scaling of (a) $C_{5}$, $C_{6}$, $C_{8}$
          for the $^{1}\Pi_{g}$-$^{3}\Pi_{u}$ pair,
          and (b) $R_{e}$ (left scale) and $D_{e}$ (right scale), 
          as a function of $n$.  
          }
\label{fig3}
\end{figure}

In Fig.~\ref{fig3}, we illustrate the scaling of the
dispersion coefficients, equilibrium distance, and well
depth of the $^{1}\Pi_{g}$-$^{3}\Pi_{u}$ pair as a function
of $n$.
In atomic units, the coefficients are given approximately by
$C_{5}\sim 3\; n^{8}$, $C_{6}\sim -0.7\; n^{11}$,
and $C_{8}\sim -50\; n^{15}$ (see Fig.~\ref{fig3}(a)). 
Neglecting $C_{8}$
in Eq.~(\ref{eq:expansion}), and setting the derivative
of $V(R)$ to zero, we find that the equilibrium distance
scales as $R_{e}\sim 0.3\; n^{3}$ $a_{0}$, in good agreement
with the numerical values shown in Fig.~\ref{fig3}(b). The scaling
of $D_{e}$ is $10^{8}n^{-7}$ cm$^{-1}$, as expected 
(see Fig.~\ref{fig3}(c)).
Note that other pairs may have different $R_{e}$ and $D_{e}$ scaling,
depending on the relative magnitude of the dispersion
terms. The wells for the $^{1}\Pi_{g}$-$^{3}\Pi_{u}$ 
pair, although shallow, 
support many vibrational bound levels. In Table~\ref{tab1}, 
we list the two lowest and highest
levels found for $n=20$, 40 and 70 \cite{note:levels}, which
support 143, 125 and 107 bound levels, respectively.
While the wells for high $n$ are much shallower than those with
smaller $n$ (e.g., see $n=20$ and 70 in Table~\ref{tab1}),
their larger extension leads to denser energy levels, and hence
they also support a large number of levels.

When using the expression for $V(R)$, the Born--Oppenheimer
approximation is assumed valid. To verify that it is the case,
we compare the vibrational period $\tau_{\rm vib}(v)$ 
of a given bound level $v$ with the typical time for the 
electron motion $\tau_{e}=h/|E_{n}|=(4\pi/\alpha c)n^{2}a_{0}$ 
where $E_{n}$ is the electron binding energy.
If $\tau_{e}/\tau_{\rm vib}\ll 1$, the Rydberg electrons of both
atoms easily follow the motion of the ions, and
the Born--Oppenheimer curve $V(R)$ needs no further diabatic
corrections. Taking the levels in Table~\ref{tab1} as examples,
for $n=20$, we find $\tau_{e}/\tau_{\rm vib} \sim 1.5\times 10^{-6}$
for the lowest level $v=0$, and $2.2\times 10^{-11}$ for the
highest level $v=142$: the electrons follow adiabatically the
ions for all vibrational levels. For $n=70$, the ratio becomes
$6.8\times 10^{-9}$ for $v=0$ and $4.5\times 10^{-14}$ for $v=106$. 
Naturally, the Born--Oppenheimer approximation gets better for 
the higher vibrational levels,
since they are more extended and therefore take more time to
make a full oscillation. However, the spontaneous
decay of one of the excited Rydberg atoms will limit 
the lifetime of these long range molecules and prevent the
existence of the upper lying vibrational bound levels. 
Although the lifetime of the excited atoms is long, scaling as 
$\tau_{\rm at} \sim \tau_{0}n^{3}$, where 
$\tau_{0}\sim 1.4$ ns \cite{gallagher},
it is much shorter than the vibrational period for high $v$.
E.g., for $n=20$, we find $\tau_{\rm at}\sim 11.2$ $\mu$s
and $\tau_{\rm vib}=0.083$ $\mu$s and 5.58 ms
for $v=0$ and 142, respectively, and for $n=70$, 
$\tau_{\rm at}\sim 480$ $\mu$s and 
$\tau_{\rm vib}=220$ $\mu$s and 32.8 s
for $v=0$ and 106, respectively. In effect, the upper vibrational
levels can be considered as quasi-continuum states since 
only a fraction of an entire oscillation will take place
before de-excitation.

\begin{table}
\caption{Sample of vibrational bound levels for the
$^{1}\Pi_{g}$-$^{3}\Pi_{u}$ pair.
The top corresponds to $n=20$, with
$R_{e}=2509\;a_{0}$, $D_{e}=3.740\times 10^{-2}\;\text{cm}^{-1}$,
the middle to $n=40$, with $R_{e}=20\,141\;a_{0}$,
$D_{e}=4.060\times 10^{-4}\;\text{cm}^{-1}$,
and the bottom to $n=70$, with $R_{e}=103\,380\;a_{0}$,
$D_{e}=1.042\times 10^{-5}\;\text{cm}^{-1}$.}
\label{tab1}
\begin{ruledtabular}
\begin{tabular}{@{\extracolsep{\fill}}rrrrrr}
$v$ & \multicolumn{1}{c}{$E_{b}$} & 
\multicolumn{1}{c}{$E(v)$} & \multicolumn{1}{c}{$R_{1}$} & 
$R_{2}$ &$\tau_{\text{vib}}$ \\
    & \multicolumn{1}{c}{(cm$^{-1}$)} & \multicolumn{1}{c}{(cm$^{-1}$)} & 
      \multicolumn{1}{c}{($a_{0}$)} &
($a_{0}$) & (s) \\
\hline
0   & $-3.70 \,[-2]$ & $4.01 \,[-4]$ &   $2\,454$ &   $2\,571$ & $8.3\,[-8]$ \\
1   & $-3.62 \,[-2]$ & $1.19 \,[-3]$ &   $2\,417$ &   $2\,623$ & $8.4\,[-8]$ \\
$\vdots$ &        &            &         &         &  \\
141 & $-3.46 \,[-8]$ & $3.74 \,[-2]$ &   $2\,142$ &  $49\,859$ & $1.2\,[-3]$ \\
142 & $-3.89 \,[-9]$ & $3.74 \,[-2]$ &   $2\,142$ &  $77\,220$ & $5.6\,[-3]$ \\
\hline
0   & $-4.01 \,[-4]$  & $5.04 \,[-6]$ &  $19\,651$ & $20\,701$ & $6.6\,[-6]$ \\
1   & $-3.91 \,[-4]$  & $1.50 \,[-5]$ &  $19\,331$ & $21\,163$ & $6.7\,[-6]$ \\
 $\vdots$ &        &            &         &        & \\
123 & $-3.27 \,[-10]$ & $4.06 \,[-4]$ &  $17\,111$ & $431\,262$ & 0.1 \\
124 & $-1.91 \,[-11]$ & $4.06 \,[-4]$ &  $17\,111$ & $762\,169$ & 0.8 \\
\hline
0   & $-1.03 \,[-5]$ & $1.52 \,[-7]$ & $100\,574$ & $106\,629$ & $2.2\,[-4]$ \\
1   & $-9.97 \,[-6]$ & $4.51 \,[-7]$ &  $98\,762$ & $109\,334$ & $2.3\,[-4]$ \\
 $\vdots$ &        &            &         &        & \\
105 & $-1.02 \,[-11]$ & $1.04 \,[-5]$ & $87\,306$ & $2\,211\,658$ & 3.1\\
106 & $-3.57 \,[-13]$ & $1.04 \,[-5]$ & $87\,306$ & $4\,337\,482$ & 32.8\\
\end{tabular}
\end{ruledtabular}
\end{table}

Another effect limiting the existence of Rydberg long range
molecules is auto-ionization. When two excited atoms interact,
one atom can decay to a lower excited state while the second
atom is ionized, the free electron picking up most of the 
kinetic energy. However, if the separation between the Rydberg 
atoms is larger than $R_{\rm LR}$,
there is little overlap between their
electronic clouds, and one expects the auto-ionization probability 
to be small (the atoms interact via their electric dipoles at such 
distances). 
Note also that for alkali dimers in general \cite{francoise}, 
and Rb in particular, avoided crossings with the ionic curves (here
correlated to Rb$^{+}+$Rb$^{-}$) perturb the 
lowest asymptotes
such as $5p-5p$. However, for the case where the Le~Roy radius condition
is satisfied, i.e., for $n\geq 20$ (see Fig.~\ref{fig2}(b)), the
$np-np$ asymptotes are much higher than the ionic curves, and no
perturbation from avoided crossings will occur.

\begin{figure}
\centerline{\epsfxsize=3.25in\epsfbox{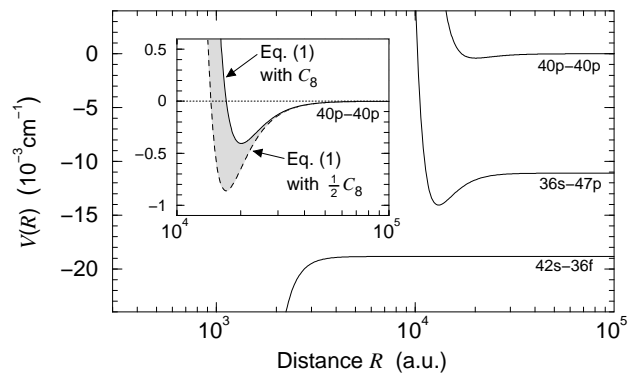}}
\caption{ Nearest asymptotes to $40p-40p$ for the 
          $^{1}\Pi_{g}$-$^{3}\Pi_{u}$ symmetries. Inset:
          convergence of Eq.(\ref{eq:expansion}) for $n=40$. 
          The shaded area gives an estimate of
          the truncation error.}
\label{fig4}
\end{figure}

The density of electronic states in the Rydberg region is very large, and
avoided crossings from potential curves correlated to other asymptotes
with the same symmetry
could perturb the shallow wells. However, the 
closest asymptotes are generally separated by several times the potential depth
$D_{e}$, and no avoided crossing in the well region 
occur \cite{note:degenerate}. Fig.~\ref{fig4} illustrates
the well for $40p-40p$ and the two nearest curves correlated to
$36s-47p$ and $42s-36f$, for the $^{1}\Pi_{g}$-$^{3}\Pi_{u}$ symmetries.
The separation between them
is much larger than $D_{e}$ and no crossings occur. 
Although the various asymptotes become denser as $n$ increases,
the potentials become very shallow ($D_{e}\propto n^{-7}$), and
generally no crossing is found. Note that avoided crossings from
a lower asymptote repulsive curve and a higher asymptote attractive
curve (with the same symmetry) could, in fact, produce deep wells.

For these extremely extended states, retardation
effects may become important. Again, to estimate the importance
of the photon time-of-flight, we compare the electronic time
$\tau_{e}$ with the time it takes a photon to cover the distance
between the two centers: $\tau_{\rm ph}\sim R/c$. For the highest
levels, the relevant distance is the outer turning point $R_{2}$, but
for the deeper levels, a good estimate is obtained by using the
equilibrium distance $R_{e}$. Using the scaling $R_{e}\sim 0.3\; n^{3}$,
we have $\tau_{\rm ph}/\tau_{e}\sim (0.3\alpha/4\pi)n\ll 1$ for $n<100$.
For the highest vibrational levels $v$ in 
Table~\ref{tab1}, we get $\tau_{\rm ph}/\tau_{e}\sim 0.06, 0.07,$ and
0.11, for $R_{2}$ of $v=140, 141,$ and 142, respectively (for $n=20$),
and $\tau_{\rm ph}/\tau_{e}\sim 0.19, 0.26,$ and 0.51 for 
$R_{2}$ of $v=104, 105,$ and 106, respectively (for $n=70$). 
Although retardation effects are not important for the 
lower vibrational levels (with $R_{2}$ slightly larger than $R_{e}$),
even for large $n$, one needs to consider them for the higher 
vibrational levels. The sensitivity of the interaction potential
to retardation effects could actually be used to detect these 
effects with high accuracy using high precision spectroscopy.

We included three terms in the asymptotic $1/R$ expansion 
(\ref{eq:expansion}), but one has to verify the effect of the
truncation on the well properties. The standard procedure
to estimate the maximum error due to the truncation of
higher terms is to include half of the last term \cite{alex}: 
here, $C_{8}$.
In Fig.~\ref{fig4}, we illustrate the effect of such variation 
for $n=40$. The well becomes deeper, but is still
located outside the Le~Roy radius, hence the existence of long 
range wells is robust to the truncation of the $1/R$ expansion.

To detect these Rydberg macrodimers, one could photoassociate
them from the ground state ($5s$ for Rb) to the target
molecular symmetry, using appropriate photon wavelength and
polarization.
After the photoassociation pulse
\cite{note:photoassociation},
one can field-ionize the system with an electric pulse and 
detect the ions. The existence
of the molecules should lead to small peaks red-detuned from
the atomic asymptote. Because the spacing of the bound levels is
small, individual levels will be hard to detect (a laser 
width smaller than the spacing $\Delta$ is needed: for $n=20$, 
$\Delta\sim 10^{-3}$ cm$^{-1}\sim 300$ MHz for the deepest levels),
and the most probable signature would be an overall red-detuned
peak. The wells are shallow, and only those deep enough could 
be detected. To enhance the sensitivity of the detection scheme,
one could also use radio-frequency to excite a bound level to a
continuum state corresponding to another asymptote (e.g., to 
$(n+1)s-np$). The detection of any $(n+1)s$ atoms (e.g., by 
ramp-field ionization) would be attributed to
the existence of a long range bound molecular state, since the
RF field would not be resonant with free $np$ atoms.

We have computed long range interactions between two Rb
atoms in the same $np$ state, and found that shallow long range 
wells supporting several vibrational levels
exist for certain molecular symmetries.
Although specific calculations were
performed for Rb and $np-np$ asymptotes, 
the existence of these macrodimers is general: one can expect them
for various asymptotes $n_{1}\ell_{1}-n_{2}\ell_{2}$ and
for all alkali atoms.
Furthermore, avoided crossing between curves of the same 
symmetry with different asymptotes could also provide deep
long range molecular wells, in a manner similar to the long 
range wells observed in many alkali dimers \cite{dimers}.
The macrodimers are extremely sensitive to their
environment, and as such, they could be used as probes for
extremely weak interactions, e.g., to
measure retardation effects and vacuum fluctuations, or
any weak electromagnetic interaction. Also, due to
their very small rovibrational energy splittings, 
macrodimers would provide a unique tool to study 
quenching in ultracold collisions,
since they would remain trapped (as
opposed to usual dimers \cite{bala}). Finally,
the detection of such
exotic molecules in itself would be a considerable achievement.

\acknowledgements{
The authors thank A. Dalgarno, V. Kharchenko,
P.L. Gould, and E. Eyler for helpful discussions. This work 
was supported by the University of Connecticut Research Foundation,
the Research Corporation, and the Grant ITR-0082913 from the
National Science Foundation.}


\begin{thebibliography}{99}
\bibitem{note:photoassociation} A short intense laser pulse 
  (e.g., 10MW/cm$^{2}$ during 10 ns) photoassociates all atom pairs with the 
  right separation. The number of macrodimers produced will depend
  only on the pairs' spatial distribution.
\bibitem{note:qd} The dipole, quadrupole, and octupole radial matrix
  elements were obtained using the quantum defect of $p_{1/2}$
  to evaluate electronic wavefunctions. See 
  {\protect\cite{gallagher}}.
\bibitem{note:degenerate} For rare accidental near-degeneracies  
  where other asymptotes are close compared to $D_{e}$, one needs to
  diagonalize the appropriate interaction subspace.
\end{thebibliography}
\end{document}